\documentclass{jpsj-suppl}
\usepackage{times}
\usepackage{color}
\usepackage{ulem}

\title{
Phonons and Spin Excitations in Fe-Based Superconductor
Ca$_{10}$Pt$_4$As$_8$ (Fe$_{1-x}$Pt$_x$As)$_{10}$ ($x \sim 0.2$)
}

\author{Kazuhiko IKEUCHI$^{1}$
\thanks{E-mail address: k{\_}ikeuchi@cross.or.jp},
Masatoshi SATO$^{1}$,
Ryoichi KAJIMOTO$^{1, 2}$,
\\
Yoshiaski KOBAYASHI$^{3}$,
Kazunori SUZUKI$^{3}$,
Masayuki ITOH$^{3}$,
Philippe BOURGES$^{4}$,
Andrew D. CHRISTIANSON$^{5}$,
Hiroki NAKAMURA$^{6}$,
Masahiko MACHIDA$^{6}$
}
\inst{
$^{1}$Research Center for Neutron Science and Technology, Comprehensive Research Organization for Science and Society, Tokai, Ibaraki 319-1106, Japan\\
$^{2}$Materials and Life Science Division (MLF),J-PARC Center, Tokai, Ibaraki 319-1195, Japan\\
$^{3}$Dept. of Phys., Nagoya Univ., Furo-cho, Nagoya 464-8602, Japan\\
$^{4}$Lab. L$\acute{e}$on Brillouin, CEA-CNRS, CEA Saclay, 91191 Gif-sur-Yvette Cedex, France\\
$^{5}$Quantum Condensed Matter Division, ORNL, Oak Ridge, TN 37831, USA\\
$^{6}$CCSE, Japan Atomic Energy Agency, 5-1-5 Kashiwanoha, Kashiwa, Chiba, 277-8587, Japan
}
\abst{
By means of neutron inelastic scattering, magnetic excitations and phonons were measured for a single crystal of slightly overdoped superconductor Ca$_{10}$Pt$_4$As$_8$ (Fe$_{1-x}$Pt$_x$As)$_{10}$ ($x \sim 0.2$) with the transition temperature $T_{\rm c}$ of $\sim$33 K.
Below $T_{\rm c}$, magnetic excitation spectra $\chi "(\mbox{\boldmath $Q$}, \omega)$ measured at \mbox{\boldmath $Q$} = \mbox{\boldmath $Q_{\rm M}$} [magnetic $\Gamma$ points] are gapped, and in the relatively higher $\omega$ region, the $\chi "(\mbox{\boldmath $Q_{\rm M}$}, \omega)$-increase was observed with decreasing $T$, where the maximum of the increase was found at $\sim$18 meV at 3 K ($<< T_{\rm c}$).
These characteristics are favorable to orbital-fluctuation-mediated superconductivity with the so-called $S_{++}$ symmetry of the order parameter. The energy dependence of $\delta$\mbox{\boldmath $Q$}-temperature($T$) curve of the magnetic excitations seems to have anomalous behavior in rather wide $T$ region above $T_{\rm c}$, $\delta$\mbox{\boldmath $Q$} being the width of the \mbox{\boldmath $Q$} scan profile.
In the phonon measurements, we observed softening of the in-plane TA mode, which corresponds to the elastic constant $C_{66}$. This softening seems to start at rather high temperature $T$, as $T$ is lowered. Additionally, anomalous increase in spectral weights of the TO-phonons at around \mbox{\boldmath $Q_{\rm M}$} in the region $35 < \omega < 40$ meV was found even above $T_{\rm c}$, as $T$ is lowered from ambient $T$. Because the spectral weights in this $\omega$ region mainly correspond to the in-plane motions of Fe atoms and because orbital fluctuations are expected to be strong at around \mbox{\boldmath $Q_{\rm M}$}, the result may present clues to investigate a possible coupling between the fluctuations of the orbitals and lattice system.
}

\kword{Fe-doped superconductor, neutron inelastic scattering, phonons, magnetic excitations}

\begin{document}
\maketitle

\section{Introduction}

To understand high critical temperature of superconducting transition in Fe-based systems, it is important to study detailed normal state properties of conducting FeAs layers.
Although it is widely believed that the superconducting pair formation is due to the spin-fluctuation exchange, which should generate sign reversal between the superconducting order parameters $\Delta$ on the Fermi surfaces around $\Gamma$ and M points [magnetic $\Gamma$ points] of the (pseudo) tetragonal lattice in the reciprocal space ($S_{\pm}$ symmetry), roles of orbital fluctuations (occupancy fluctuations among the 3$d$ orbitals) have also been proposed as an alternative origin of the pair formation [1-4].
If the mechanism via orbital fluctuations, which characterizes multi band systems, is relevant, the order parameter $\Delta$ does not have the sign reversal ($S_{++}$ symmetry).
One of striking differences between above two symmetries can be expected in effects of non-magnetic impurities on $T_{\rm c}$: Although non-magnetic impurities do not suppress, as is well-known, $T_{\rm c}$ values for systems without sign change of $\Delta$ over the Fermi surface(s), rapid suppression of $T_{\rm c}$ values is expected for systems with sign change of $\Delta$.
Actually, it has been shown that rates of $T_{\rm c}$-suppression by non-magnetic impurities in the conducting planes are too small to be explained by the impurity scattering for the $S_{\pm}$ symmetry [5, 6].
After taking the result into account, we can conclude that impurities doped into the FeAs planes do not act as the pair breaking centers, indicating that the $S_{\pm}$ symmetry is not likely.
Then, we have to be careful not to miss a possible novel superconducting mechanism, even though the spin fluctuation mechanism is widely considered to be the most plausible one.

In order to discuss which one of spin and orbital fluctuations is primarily important in Fe-based systems, $T$-$x$ phase diagrams provide important information.
For example, in the case of Ba(Fe$_{1-x}$Co$_x$)$_2$As$_2$, there are characteristic temperatures $T_{\rm N}$ and $T_{\rm s}$ ($>T_{\rm N}$) of antiferromagnetic and tetragonal$\rightarrow$orthorhombic transitions, respectively [7].
It also has the onset temperature $T^{\ast}$ ($>T_{\rm s}$) of the gradual change to the so-called ``nematic'' state, which is characterized by breaking of the $C_4$ symmetry of various static quantities with decreasing $T$ (even in the tetragonal phase) [8, 9].
From the relations $T_{\rm N} < T_{\rm s} < T^{\ast}$, we presume that the observed structural transition is related to ordering of $3d_{yz}$ and $3d_{zx}$ orbitals and that the $C_4$ symmetry breaking can be understood as a kind of precursor of this ordering, because the fluctuation (electron-occupancy fluctuation) between these orbitals or their ordering is naturally expected to couple to the orthorhombic distortion.
On this point, it has been pointed out theoretically [2, 3] that the orbital fluctuation is strong at around both $\Gamma$ and M points of the (pseudo) tetragonal lattice in the reciprocal space, implying that evidence for the strong effects of the fluctuations on dynamical properties of lattice system can be found experimentally.
Thus, we consider that in the phase diagram all the spins, orbitals, lattices, and couplings among these freedoms are involved as important ingredients.
The question is which one of them has a primary role in the determination of the phase diagram.

To get consistent understanding of these ingredients, we have performed inelastic neutron scattering studies on magnetic excitations and phonons using a large crystal sample ($\sim$ 10 g) of Ca$_{10}$Pt$_4$As$_8$ (Fe$_{1-x}$Pt$_x$As)$_{10}$ prepared at Nagoya University.
The system belongs to one of two series of Ca$_{10}$Pt$_n$As$_8$ (Fe$_{1-x}$Pt$_x$As)$_{10}$ systems with $n$ = 3 (10-3-8 phase) and 4 (10-4-8 phase) found by Kakiya et al. [10, 11].
Below, we present a brief report on the data obtained by varying $T$ through $T_{\rm c}$, which may indicate possible evidence for orbital-fluctuation-mediated superconductivity.

\section{Experiment}

\begin{figure}
\begin{center}
\includegraphics[width=15cm]{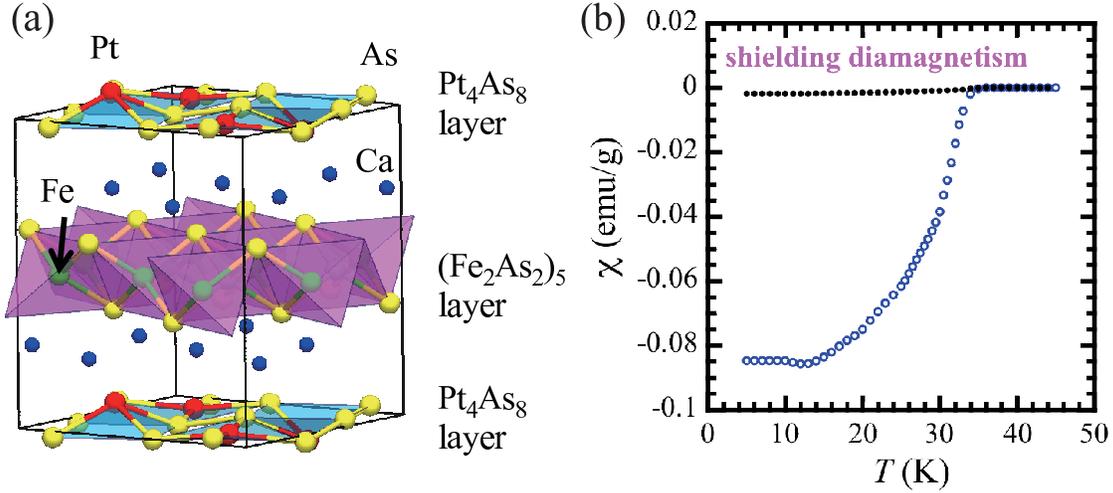}
\end{center}
\caption{
(a) Crystal structure of Ca$_{10}$Pt$_4$As$_8$ (Fe$_{1-x}$Pt$_x$As)$_{10}$. (b) Temperature dependence of magnetic susceptibility. The black and blue circles show the results in field and zero-field cooling conditions, respectively, where the field was parallel to the $c$ direction.
}
\label{f1}
\end{figure}

The structure of Ca$_{10}$Pt$_4$As$_8$ (Fe$_{1-x}$Pt$_x$As)$_{10}$ is rather complicated [11, 12].
It has  Fe$_{1-x}$Pt$_x$As planes common to other Fe-based superconductors and PtAs$_2$ planes, too [see Fig. 1(a)].
For this system, we use a pseudo tetragonal cell with the lattice parameters $a$ = $b$ = 3.9029(3) Å, and $c$ = 10.5122(6) \AA.
This pseudo tetragonal unit cell has only one Fe$_{1-x}$Pt$_x$As conducting layer with two Fe$_{1-x}$Pt$_x$ sites in a cell.
The shielding diamagnetism taken by using a conventional SQUID magnetometer for a small piece of the present crystal has the onset $T_{\rm c}$ value of $\sim$ 33 K [see Fig. 1(b)].
The $x$ value determined by the intensity analysis of NMR spectra [6, 13] is $\sim$ 0.20. These $x$ and $T_{\rm c}$ values are consistent with the $T$-$x$ phase diagram [10].
The PtAs$_2$ layers do not have important roles in determining the transport and magnetic properties, as will be reported elsewhere [13].
Although the fraction ($\sim$ 0.2) of Pt impurities at Fe sites in FeAs planes is large, the optimal superconducting transition temperature is very high, $T_{\rm c}$ $\sim$ 38 K, indicating that the high-$T_{\rm c}$ superconductivity of this system is robust against impurity scattering.

We carried out X-ray powder diffraction measurements on pulverized crystallites of the specimen and confirmed a little amount of the 10-3-8 phase, which does not significantly affect our results.
It should be added that the sample used in this study had two domains with similar volumes.
It was confirmed by neutron measurements of the (2, 0, 0) Bragg reflections. The peak width FWHM of each domain was $\sim$ 1.3$^{\circ}$, and they were $\sim$ 2.5$^{\circ}$ away from each other.
Thus the signals of each domain were not contaminated with that of other domain at least around the (2, 0, 0) reflection, where the in-plane TA phonon measurement was carried out. Thus the effective sample amount used here is $\sim$ 5 g.

Neutron data of the spin excitations and low-energy phonons were collected with both thermal (2T) and cold (4F2) triple-axis-spectrometers (TAS) at the neutron reactor ORPHEE of Laboratoire Leon Brillouin (LLB), France, respectively.
Incident and scattered beams were focused by pyrolytic graphite 0 0 2 monochrometer and analyzer, respectively, and in these cases, collimation conditions were fully open.
Neutron data of optic phonons were collected with thermal TAS (HB-3) at the neutron reactor HFIR of Oak Ridge National Laboratory, USA.
Collimation condition was 48'-40'-40'-70'.
In all cases, pyrolytic graphite filter was placed before the analyzer to eliminate the higher order reflections.

\section{Results and Discussion}
\subsection{
 $T$ dependence of the dynamical magnetic susceptibility across $T_{\rm c}$
}

\begin{figure}
\begin{center}
\includegraphics[width=15.5cm]{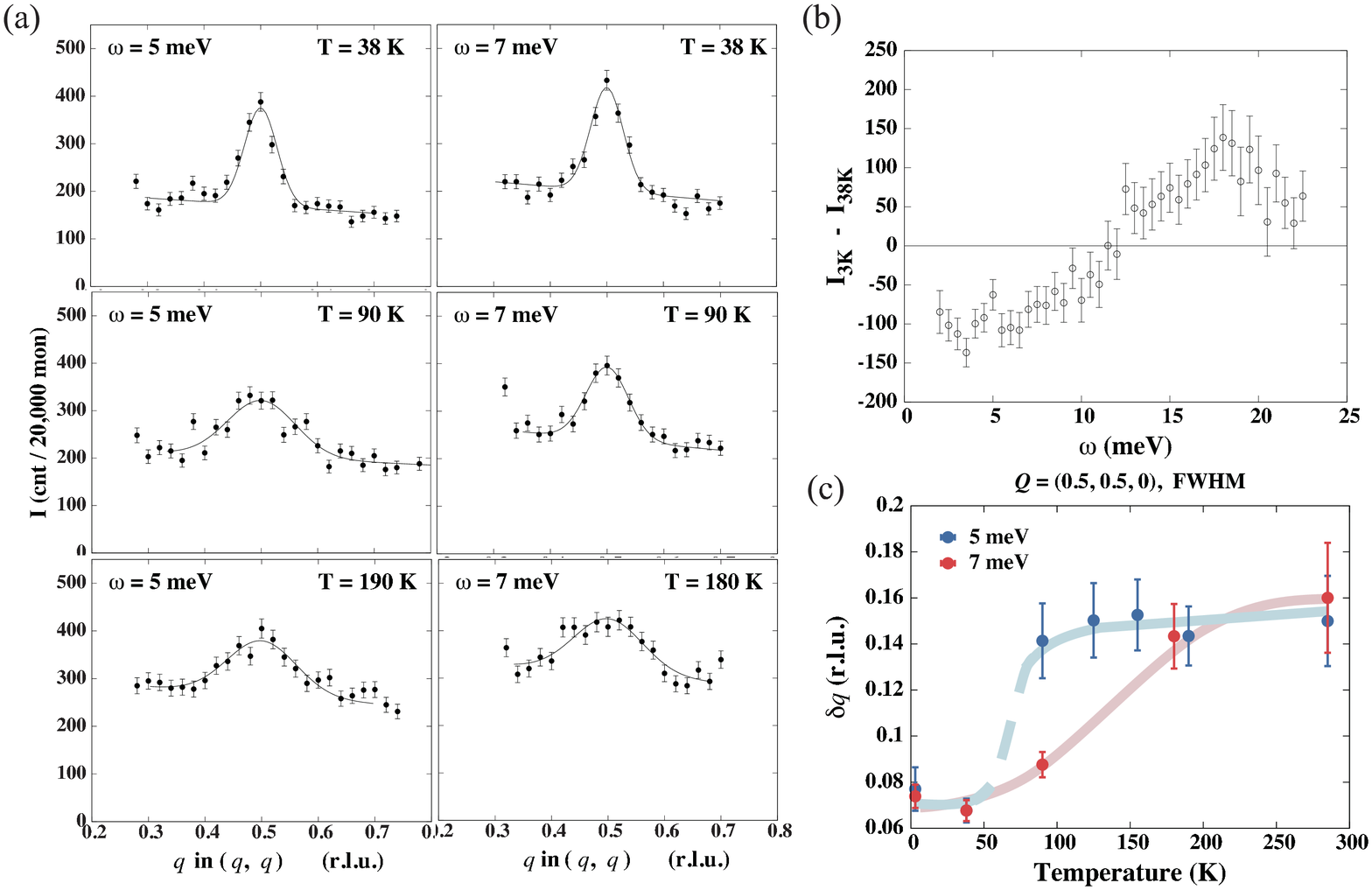}
\end{center}
\caption{
(a) \mbox{\boldmath $Q$}-scan profiles of the magnetic excitation at an M point (0.5, 0.5, 0) or (0.5, 0.5) in the 2-dimentional notation at various temperatures, measured with $\omega$ = 5 meV and 7 meV.
(b) The difference of the spectral weights of the magnetic excitation between $T$ = 5 K and 38 K. These data were obtained at an M point (0.5, 0.5, 2).
(c) $T$ dependence of the \mbox{\boldmath $Q$}-scan widths of the magnetic peaks along ($q$, $q$, 0) through (0.5, 0.5, 0) with $\omega$ = 5 meV (blue circle) and 7 meV (red circle).
All the data were taken with $E_{\rm f} = 14.6$ meV.
The colored lines are guide to the eyes.
}
\label{f1}
\end{figure}

As examples of our data series, Figure 2(a) shows the peak profiles of the magnetic scattering intensity $S(\mbox{\boldmath $Q$}, \omega) \propto (n+1)\cdot \chi "(\mbox{\boldmath $Q$}, \omega)$ at selected $T$ values, where $n$ is the Bose factor.
They were obtained above $T_{\rm c}$ by scanning $Q$ at fixed $\omega$ values of 5 and 7 meV along ($q$, $q$, 0) through an M point (1/2, 1/2, 0).
We can see that the peaks are always located at \mbox{\boldmath $Q_{\rm M}$} in the measured $\omega$ region less than 11 meV by constant-$\omega$ scans, indicating that the dispersion curve of the magnetic excitation is very steep in the \mbox{\boldmath $Q$}-$\omega$ plane, as was already reported for one of other 10-4-8 crystals by using the chopper spectrometer [14].

Figure 2(b) shows the difference between the magnetic excitation spectra observed at $T$ = 3 K and 38 K by scanning $\omega$ at an M point (1/2, 1/2, 2).
With the development of the superconducting gap on cooling through $T_{\rm c}$, the spectral weight shifts from the low-energy region to the high-energy region, and we can see that the maximum increase of the spectral weight is at the energy $\omega = E_{\rm p} \sim$ 18 meV.
If the peak is due to the coherence factor effect characterizing the superconducting condensed state with the $S_{\pm}$ symmetry, the peak position $E_{\rm p}$ is expected to be smaller than the value $|\Delta_{\Gamma}|$+$|\Delta_{M}|$, where $\Delta_{\Gamma}$ and $\Delta_{M}$ are superconducting order parameters on the Fermi surfaces around $\Gamma$ and M points, respectively[15].
Here, judging from the experimental values of the $\Delta$ values observed for various systems, the observed value of $E_{\rm p}/k_{\rm B}T_{\rm c}$ ($>$ 6) can hardly be considered to satisfy the condition $E_{\rm p} <$ $|\Delta_{\Gamma}|$+$|\Delta_{M}|$, indicating that the peak is not due to the effect of coherence factor of the $S_{\pm}$ symmetry, but probably due to a different effect predicted for the $S_{++}$ symmetry of $\Delta$ by Onari et al.[16].
Additionally, we have found the rather large difference of the \mbox{\boldmath $Q$}-scan profile widths ($\delta$\mbox{\boldmath $Q$}) of the magnetic excitation between $\omega$ = 5 meV and 7 meV from the data at $T$ = 90 K.
The $\delta$\mbox{\boldmath $Q$} with $\omega$ = 5 meV is broader than one with $\omega$ = 7 meV.
$T$ dependences of $\delta$\mbox{\boldmath $Q$} are plotted in Fig. 2 (c).
We can say that, in the $T$ region of 38 K $< T <$ 180 K, the $\delta$\mbox{\boldmath $Q$}-$T$ curve exhibits anomalous $\omega$ dependence.
It is interesting to see at what temperature the anomaly occurs with decreasing $T$, because it may be related to the appearance of the nematic phase stated in the introduction.

\subsection{Temperature dependence of phonons}

\begin{figure}
\begin{center}
\includegraphics[width=15cm]{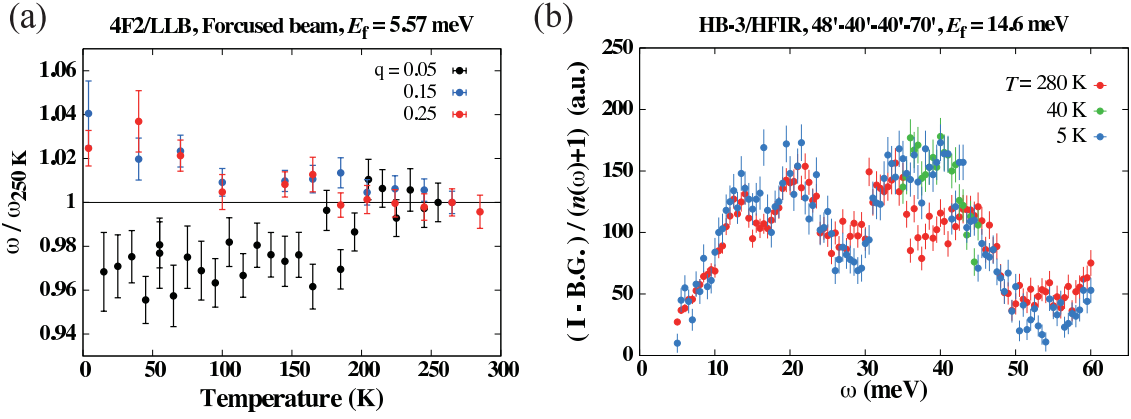}
\end{center}
\caption{
(a) Frequencies of in-plane TA mode of Ca$_{10}$Pt$_4$As$_8$ (Fe$_{1-x}$Pt$_x$As)$_{10}$ taken at \mbox{\boldmath $Q$} = (2, $q$, 0) are plotted against $T$.
Each frequency is scaled by the value at $T$ = 250 K.
(b) Energy spectrum at a fixed wave vector of \mbox{\boldmath $Q$} = (2.44, 0.49, 3), close to an M point (2.5, 0.5, 3).
These spectra are corrected by Bose factor after subtraction of constant background estimated with high energy region.
}
\label{f1}
\end{figure}

First, we present result of measurements of the in-plane transverse acoustic mode, which corresponds to the elastic constant $C_{66}$.
The $\omega$-scan profiles were taken at various fixed $q$ values of \mbox{\boldmath $Q$} = (2, $q$, 0) and at various $T$ points between 4 K and room temperature to obtain the $q$-$\omega$ dispersion relation.
Figure 3(a) shows the $T$ dependence of the frequencies at the transverse propagating vector $q$ = 0.05, 0.15, and 0.25.
Each frequency is normalized by the one at $T$ = 250 K.
The softening is observed at $q$ = 0.05 with decreasing $T$, while hardenings are observed at the other $q$ positions, which are away from the $\Gamma$ point $q$ = 0.
The softening amplitude at $q$ = 0.05 can be estimated to be about 8 \% with considering the correction of the hardening observed with decreasing $T$ in the other $q$ positions in the figure.
Reducing the softening amplitude into a value of the elastic constant, it becomes nearly 16 \%, which is similar order of magnitude to that of the Ba122 in the superconducting phase[17].
Therefore, the result suggests that the coupling of the lattice system seems to commonly exist in Fe-based superconductors.

Figure 3(b) shows temperature variation of energy spectra of the phonons at \mbox{\boldmath $Q$} = (2.44, 0.49, 3.0) near an M point \mbox{\boldmath $Q$} = (2.5, 0.5, 3), where the in-plane optical phonon modes are observed.
The enhancement of the intensity was observed at around ω = 40 meV at low temperature region.
It sets in at temperature higher than $T_{\rm c}$ = 33 K, at least higher than 40 K, with decreasing $T$.
Comparing the phonon spectra with the first principle calculation of phonon density of states, we think the strong enhancement arises from anomalous behavior of the in-plane motion of Fe atoms, suggesting strong coupling between the orbital fluctuations and lattice system at around \mbox{\boldmath $Q_{\rm M}$}.

\section{Summary}

We have performed measurements of spin and lattice systems of Ca$_{10}$Pt$_4$As$_8$ (Fe$_{1-x}$Pt$_x$As)$_{10}$ by inelastic neutron scattering to identify the microscopic origin of the superconductivity.
Even though there exist a strong idea that the spin-fluctuation is relevant to the occurrence of the superconductivity, we should consider the orbital fluctuation mechanism in order to explain the result that the broad peak structure in the magnetic excitation spectra appears near $\omega$ = $E_{\rm p}$ = 18 meV ($E_{\rm p}/k_{\rm B}T_{\rm c} \geq$ 6) at low temperatures $T << T_{\rm c}$, which is favorable to the $S_{++}$ symmetry of $\Delta$.
Actually, we can succeed in observing the softening of the acoustic phonon at $\Gamma$ point and the anomalous increase of the spectral weight of in-plane optical phonons at the M point with decreasing $T$, suggesting the strong coupling between orbital fluctuations and lattice system.

As described above, there are experimental suggestions that the orbital fluctuations are actually playing an essential role in the realization of the superconductivity in Fe-based system, and it is informative to study normal state properties, in which all the spins, orbitals and their coupling to the lattice system are involved.

\begin{flushleft}
\large{\bf{Acknowledgment}}\\[4mm]
\end{flushleft}

We thank Prof. H. Kontani for fruitful discussion. They also thank many other collaborators at Nagoya University, JAEA and Tokyo University of Science. The work is supported by Grants-in-Aid for Scientific Research from the Japan Society for the Promotion of Science (JSPS) and Technology and JST, TRIP. The experiments at ORFEE at Saclay was carried out by a project No.10604.
The measurements at HFIR of ORNL were carried out by a project No. ITPS-7059 with the supports of the travelling expense by U.S.–Japan Cooperative Program on Neutron Scattering Research.
Research at HFIR of ORNL was sponsored by the Division of Scientific User Facilities of the Office of Basic Energy Sciences, US Department of Energy.

\end{document}